\documentclass[structabstract]{aa}
\usepackage{txfonts}
\usepackage{natbib}
\bibpunct{(}{)}{,}{a}{}{,} 
\usepackage{graphicx}

\begin{document}
\title{The distance to the C component of I Zw 18 and its star formation 
history}
\subtitle{A probabilistic approach}
\author{Luc Jamet\inst{1,2}  \and Miguel Cervi\~no\inst{1} \and Valentina Luridiana\inst{1,3} \and Enrique P\'erez\inst{1}, and T. Yakobchuk\inst{4}}

\institute{Instituto de Astrof{\'\i}sica de Andaluc{\'\i}a (CSIC). Camino bajo de Hu\'etor, 50, Granada 18080, Spain
\and
Instituto de Astronom{\'\i}a, Universidad Nacional Aut\'onoma de M\'exico, Apartado Postal 70-264, Ciudad Universitaria, M\'exico D.F. 04510, Mexico 
\and
Instituto de Astrof{\'\i}sica de Canarias (IAC). c/V{\'{\i}}a L\'actea s/n, La Laguna, E38205 Tenerife, Spain
\and
Main Astronomical Observatory, Zabolotnoho 27, Kyiv 03680, Ukraine }

\date{ Received 09 July 2007 / Accepted 26 August 2009}
\abstract {}
{We analyzed the resolved stellar population of the C component of the extremely metal-poor dwarf galaxy \object{I Zw18} in order to evaluate
its distance and star formation history as accurately as possible. In particular, we aimed at answering the question of whether this stellar
population is young.}
{We developed a probabilistic approach to analyzing high-quality photometric data obtained with the Advanced Camera for Surveys 
of the Hubble Space Telescope. This approach gives a detailed account of the various stochastic aspects of star formation. We carried out two 
successive models of the stellar population of interest, paying attention to how our assumptions could affect the results.}
{ We found a distance to the C component of I Zw 18 as high as 27 Mpc, a significantly  higher value  than those cited in previous works. 
The star formation history we inferred from the observational data shows various interesting features: a strong starburst that lasted for 
about 15 Myr, a more moderate one that occurred 
$\approx$ 100 Myr ago, a continuous process of star formation between both starbursts, and a possible 
episode of low level star formation
 at ages over 100 Myr. The stellar population studied is likely 
$\approx$ 125 Myr old, although ages of a few Gyr 
cannot be ruled out. Furthermore, nearly all the stars were formed in the last few hundreds of Myr.}
{}
\keywords{ Galaxies: individual (I Zw 18), Galaxies: photometry, Galaxies: stellar content,  Galaxies: formation 
}
\authorrunning{L. Jamet et al.}
\titlerunning{The distance and SFH of  I Zw 18 C}
\maketitle

\section{Introduction}

The study of dwarf galaxies is an important topic for the understanding
of how galaxies form and evolve. Their presence
in galaxy clusters and their physical properties put constraints
on cosmological models, especially on the dark matter content
of the Universe \citep[e.g.][]{Retal05}. Furthermore, 
because their shallow gravitational potential well,
 dwarf irregular (dIrr) galaxies are ideal
benchmarks for studing the effects of various processes, either
internal (e.g., stellar winds and supernovae) or external (like
gas infall and tidal forces), on the triggering and regulation
of star formation (SF) in galaxies \citep{HEB98,HR02}. Finally, 
metal-deficient dIrr galaxies are
expected to match the chemical composition of pristine galaxies,
and efforts have been made to derive the primordial helium
abundance of the Universe from the spectroscopic analysis
of their ionized gas 
\citep{PTP74,PTP76,ITL97, ITS07,PLP07}.

Among the dIrr galaxies known, I~Zw18 is arguably the
most fascinating one. 
This object is the second lowest
metallicity galaxy known (12 +log O / H = 7.2), corresponding
to $\approx$ 1/30 of the solar value \cite[e.g.][]{Petal92,SK93,IT98,Ietal99},
being the lowest one SBS 0335-052 (West) \citep{Ietal04}.
Its blue color \citep{vZetal98} and its intense
nebular emission 
\cite[e.g.][]{VIP98,Cetal02}
 bears witness to an
intense ongoing SF episode. The galaxy presents a complex
morphology; it is dominated by a two-lobed body (the ``main
body") 
with one companion,
 the ``C component"
\cite[hereafter \object{I Zw18C}][]{Detal96a}.

In spite of many efforts, the distance and, more important,
the age of the galaxy have not been well determined yet. By analyzing
HST/WFPC2 color-magnitude diagrams (CMDs) and
HST/FOS spectra, and assuming a distance of 10 Mpc (applying
straightly the Hubble law to the redshift of the galaxy),
\cite{Detal96b} detected a stellar population of up to 50
Myr in the main body of the galaxy and up to 300 Myr in the
C component. They found no support for or against an old underlying population. Assuming the same
distance and examining a larger set of HST/WFPC2 photometric
data, \cite{ATG99} decomposed the star formation history
(SFH) in a continuous SF lasting over $\sim$ 1 Gyr and an ongoing
starburst that started $\approx$ 15--20 Myr ago. By analyzing 
HST/WFPC2 CMDs jointly, ground-based spectra and the morphology
of some nebular features, \cite{Ietal99} find
that a distance of 20 Mpc is necessary to explain the ionization
state of the gas. They also state that the age of the galaxy,
as derived from all three kinds of data, does not exceed 100
Myr, in agreement with the conclusion from \cite{IT99} that very 
metal-deficient galaxies must all be young.
\cite{Ost00} carried out an analysis of HST/NICMOS photometric
data and detected asymptotic giant branch stars (AGB) that,
at the adopted distance of 12.6 Mpc (derived from the redshift
corrected for the Virgocentric flow), are at least 1 Gyr old. From
the surface distribution of the fluxes and colors of the galaxy,
\cite{KO00} argue that I~Zw18 is an old galaxy
whose age is likely $\sim$ 5 Gyr. A significant age of possibly several
Gyr has also been advocated by \cite{Letal00} to account
for the very homogeneous distribution of heavy elements
throughout the ionized, optical-emitting gas of the galaxy. By
studying the deep spectra of I~Zw18C, \cite{Ietal01} find
that a distance of $\approx$ 15 Mpc is 
needed to reconcile ionized gas in this object with the apparent magnitudes of its
brightest blue stars. This distance can actually be interpreted as
a lower limit, since at larger distances these stars would be still
considered as ionizing. Their spectra are well-fitted by models
with a continuous star formation extending over the past 10--100 Myr, with no evidence of any older underlying stellar population.
\cite{Petal02} compared surface brightness
maps obtained through various optical and near-infrared filters
with the HST, which did not reveal any stellar population older
than 0.5 Gyr. Such a population may exist, but most of the stellar
content must be younger than this age. \cite{IT04}
carried out deep HST/ACS observations of I~Zw18, improving
by $\approx$ 2 mag the depth of stellar photometric data with respect
to previous efforts. They concludes that the distance to I~Zw18
should lie somewhere between 12.6 and 15 Mpc with the latter
value being more likely, and claim that the absence of any
detected red giant is  proof that the galaxy is young ($\lesssim$ 500
Myr). Finally, by analyzing the same images as  \cite{IT04}, along with other deep HST/ACS data, \cite{Aetal07} detected a red giant branch (RGB), as well as a confirmed,
Cepheid star. They derived a distance of $\approx$ 18 Mpc and
argue that the presence of RBG stars shows that the galaxy is
old  \cite[at least 1 Gyr, but see also ][]{Tetal07}.

It is striking to see that similar approaches to the evaluation
of the age of I~Zw18 lead to contradictory results (e.g.,
\citealt{Detal96b} vs. \citealt{ATG99}, or \citealt{KO00} vs. \citealt{Petal02}). It turns out that the age estimates
from the authors mentioned above go from a ten
Myr to several Gyr! This might be explained at least partially by
some uncertainties in the studies summarized above, such as the
limited depth of the available data or the sensitivity of CMD
analysis to the distance and extinction adopted. For this reason,
we decided to carry out a new analysis of the deep HST/ACS
observations performed by \cite{IT04}, focusing
on I~Zw18C. The latter has been little studied because of
its faintness ($\approx$ 24 mag/$\Box \arcsec$ in the V band). However, 
there is a key difference between I Zw 18C and the main body that makes the C component more attractive for CMD studies: it is older.
\begin{enumerate}
\item There is a significant absence of nebular emission. Because it is older, the observed surface brightness in the C component is lower than in the main body and is predominantly produced by the stellar component. The surface brightness in the main body is produced by both the stellar and nebular components \cite[see, e.g. ][]{VIP98}. Also, compact clumps of ionized gas in the main component would mimic stellar sources, and the patchy ionized gas emission makes background subtraction more difficult.     
\item Isochrones (and their associated stellar luminosity functions) have a simpler structure at older (post Wolf-Rayet) ages, with well-defined structures excepting the blue loops that make a small contribution in the stellar luminosity function. In the case of the main component, there are identified Wolf-Rayet stars \citep{Letal97,Ietal97} such a presence is difficult to explain in low-metallicity environment with standard evolutionary models \cite{Betal02}, so more complex evolutionary models (like the inclusion of rotation) would be required. However, current models including rotation produce evolutionary tracks with complex structures \cite[see e.g. Fig. 3 in][]{MM05}, so isochrones becomes strongly dependent on the assumed initial rotational velocity, the possible distribution of rotational velocities, and the (still unexplored) validity of homology relations between different stellar tracks for rotating stars. 
\end{enumerate}

There is an additional reason for choosing the C component instead of the main body. In the distance obtained by \cite{Ietal01} from the integrated spectra of the galaxy the authors take the IMF sampling effects in the emission line spectrum  into account \citep[][ in its first version]{CLC00}. However, they did not consider IMF sampling effects in the analysis of the continuum \citep{CVGLMH02,CL06} from which the distance of 15 Mpc to I~Zw18C (and by extension, the main body) was obtained. It was also our intention to obtain an alternative value for the distance with a method where such sampling effects would be minimized (the CMD analysis).

For these reasons, we have decided to use study I~Zw18C instead of the main body. We had
built a new, improved CMD from the images of \cite{IT04} and analyzed it with a probabilistic approach  to evaluate the distance, the SFH, and the age of this component.

This paper is divided as follows. In Sect. \ref{sec:2}, we define the
probabilistic tools used in our approach. In Sect. \ref{sec:3}, we describe
the observational data and the theoretical isochrones on which
our work was based. In Sect. \ref{sec:4}, we present a first model of the
stellar population studied. Section \ref{sec:5} is dedicated to a revision of
the assumptions of this model, and in Sect. \ref{sec:6} we present a new
model with improved reliability. We infer the minimum
age of I~Zw18C and discuss the estimate of its distance. Finally,
we give our conclusions in Sect. \ref{sec:7}.

\section{A probabilistic approach}
\label{sec:2}

In considering a stellar population in which $N$ individual stars
are detected in a set of $n$ photometric bands,  note $m_{i,j}$
the magnitude of the $j$th star in the $i$th photometric band. We
first examine the case of a simple stellar population, i.e. a population
where all the stars were formed at the same time and in
the same conditions. The initial mass ${\cal{M}}_j$ of the $j$th star is considered
a random variable and the initial mass function (IMF)
$\phi({\cal{M}})$ is regarded as the probability density function (PDF) from
which  ${\cal{M}}_j$ is drawn. The expected magnitudes $\bar{m}_{i,j}$ of the star
depend on ${\cal{M}}_j$ and on properties of the stellar
population: the age $\tau$, the metallicity $Z$, the distance $d$ and the
interstellar extinction coefficients $A_i$. Furthermore, those magnitudes
are affected by photometric errors, which we suppose
follow Gaussian distributions of widths $\sigma_{i,j}$. Finally,
the photometric data do not usually cover the whole IMF of
the stellar population, since the faintest stars are not detected.
Consequently, if we call $p_c(m)$ the photometric completeness
function of the data (the probability a star of magnitudes $m_i$
is detected), then the observed magnitudes $m_{i,j}$ of the star $j$
are associated to the following likelihood function (LF):

\begin{equation}
L_j(\tau; Z,d,A) = \left( \prod_{i=1}^n \frac{1}{\sqrt{2\pi}\, \sigma_{i,j}}\right)\int_{{\cal{M}}_{\mathrm{low}}}^{{\cal{M}}_{\mathrm{up}}} \phi'({\cal{M}})\, \mathrm{e}^{-\chi^2_j / 2}\,\mathrm{d}{\cal{M}}
\label{eq:1}
\end{equation}

\noindent with

\begin{equation}
\phi'({\cal{M}}; \tau, Z,d,A) = \frac{\phi({\cal{M}}) p_c(\bar{m}(({\cal{M}}))}{
\int_{{\cal{M}}_{\mathrm{low}}}^{{\cal{M}}_{\mathrm{up}}(\tau)} \phi({\cal{M}}') p_c(\bar{m}({\cal{M}}')) \mathrm{d}{\cal{M}}'}
\label{eq:2}
\end{equation}

\noindent and

\begin{equation}
\chi^2_j({\cal{M}}; \tau, Z,d,A) = \sum_{i=1}^n \frac{(m_{i,j} - \bar{m}_{i,j})^2}{\sigma_{i,j}^2}.
\label{eq:3}
\end{equation}

\noindent In Eqs. \ref{eq:1} and \ref{eq:2}, the upper limit ${{\cal{M}}_{\mathrm{up}}(\tau)}$ of the integrals
is the maximum initial mass with which a $\tau$-old star still visible
today could have been born.

Assuming that the initial masses of the stars are mutually
independent, the LF $L_\mathrm{pop}$ associated to the distribution of magnitudes
of the $N$ stars is given by the product of the individual LFs $L_j$:

\begin{equation}
L_\mathrm{pop}(\tau,Z,d,A) = \prod_{j=1}^n L_j(\tau,Z,d,A).
\label{eq:4}
\end{equation}

\noindent In practice, it is preferable to work not with the LFs themselves,
but rather with their logarithms. Hence, we introduce
the following quantities:

\begin{equation}
{\cal{H}}_j = \log L_j,
\label{eq:5}
\end{equation}

\begin{equation}
{\cal{H}}_\mathrm{pop} =  \log L_\mathrm{pop} = \sum_{j=1}^n {\cal{H}}_j. 
\label{eq:6}
\end{equation}

In this section, we have limited the definition of $L_j$ to the
case of a simple stellar population. In what follows, we propose
another definition that can be applied to composite populations.

\subsection{Introducing the stellar formation history into the
entropy}

Many stellar populations  show prolonged periods of star formation,
in which case Eq. \ref{eq:1} cannot be used. It is necessary to introduce
a function ${\cal{N}}(\tau)$ that describes the SFH of the population studied.
Like their initial masses, the ages of the different stars are
supposed to be independent random variables. As a result, we define
${\cal{N}}(\tau) \, d\tau$ as the expectancy of the number of stars observable
today and whose ages are comprised between $\tau$ and $\tau + d\tau$.
Consequently, $L_j(\tau, Z, d, A)$ must be replaced by the following
LF:

\begin{equation}
L'_j({\cal{N}}; Z,d,A) = \frac{I_j({\cal{N}}; Z,d,A)}{\int {\cal{N}}(\tau) \, \mathrm{d}\tau}
\label{eq:7}
\end{equation}

\noindent with

\begin{equation}
I_j({\cal{N}}; Z,d,A) = \int {\cal{N}}(\tau) \, L_j(\tau; Z,d,A) \,\mathrm{d}\tau.
\label{eq:8}
\end{equation}

\noindent The entropy ${\cal{H}}_\mathrm{pop}$ hence become:

\begin{equation}
{\cal{H}}_\mathrm{pop} = \sum_{j=1}^n \log I_j({\cal{N}}; Z,d,A) - N \log \int {\cal{N}}(\tau) \, \mathrm{d}\tau.
\label{eq:9}
\end{equation}

For given values of $Z$, $d$, and $A$, maximizing ${\cal{H}}_\mathrm{pop}$ leads to
the best-fit shape of ${\cal{N}}(\tau)$, not to absolute values of this function.
The latter must then be normalized with the number of
stars used:

\begin{equation}
 \int {\cal{N}}(\tau) \,  \mathrm{d}\tau = N.
\label{eq:10}
\end{equation}

\noindent The mass rate $\dot{\cal{M}}(\tau)$ of star formation can be derived from
${\cal{N}}(\tau)$ by considering which part of the IMF is visible in the data
at the age $\tau$:

\begin{equation}
 \dot{\cal{M}}(\tau) = {\cal{N}}(\tau) \frac{\int_{{\cal{M}}_{\mathrm{low}}}^{{\cal{M}}_{\mathrm{up}}(0)}  {\cal{M}} \, \phi({\cal{M}}) \, \mathrm{d}{\cal{M}}}
{\int_{{\cal{M}}_{\mathrm{low}}}^{{\cal{M}}_{\mathrm{up}}(\tau)} \phi({\cal{M}}) p_c(\bar{m}({\cal{M}};\tau)) \mathrm{d}{\cal{M}}}.
\label{eq:11}
\end{equation}

In some cases, a parametric model can be provided for
the SFH. In this case,  ${\cal{N}}(\tau)$ can be fitted using a Levenberg-
Marquardt algorithm. In the case where no parametric
model is assumed for the SFH, the latter must be inferred in
a self-consistent way from the observed distribution of magnitudes
$m_{i,j}$. This can be achieved with the method developed by
\cite{Hetal99}.

\subsection{Comparison with other approaches}
\label{sec:2.2}

The variable ${\cal{H}}_\mathrm{pop}$ is an entropic tool whose maximization can be used to
analyze resolved stellar populations. Contrary to simpler estimators
(e.g., the chi-square one), it is sensitive not only to the
shape of the tested isochrones in color-magnitude diagrams,
but also to the density distribution of the stars along those
isochrones, through the relations between the age, mass and
magnitudes.
This is particularly true for rapid stellar phases,
where the magnitudes are very sensitive to the ages and initial
masses of the stars. 
Other probabilistic methods have been
developed to characterize resolved stellar populations,
especially the comparison of star counts between observed and
synthetic Monte-Carlo CMDs in different bins of the color-magnitude
space \cite[e.g.][]{Getal99}. However, the formalism
proposed in this work is based on rigorous concepts,
it is more accurate than the synthetic CMD approach. In
particular, it avoids the biases and errors caused by the binning of
data and the random content of simulated CMDs. Moreover, it
bypasses expensive Monte-Carlo computations for the search
of optimized parameters of the stellar population studied.
 In summary, ${\cal{H}}_\mathrm{pop}$ allows a study of the stellar luminosity function as a whole, with its correlations among different luminosity bins, instead of only portions of the luminosity function without correlations among bins.
 
The approach we have presented is of course not free of
drawbacks. In particular, there is no simple way to assess the
goodness-of-fit, and Monte-Carlo simulations may be required
to carry out this task (see Sect. \ref{sec:4.3}). Furthermore, like other
probabilistic methods, the results deeply depend on the input
physics, in particular the stellar evolution models and the conversion
of stellar parameters to magnitudes. Ideally, different
evolutionary tracks and model atmospheres, or empirical magnitude
calibrations, should be tested when analyzing an object,
at the cost of simplicity and time.

\section{Applications to I~Zw18C}
\label{sec:3}

\subsection{The photometric data}
\label{sec:3.1}

\begin{figure}
\includegraphics[width=8.5cm]{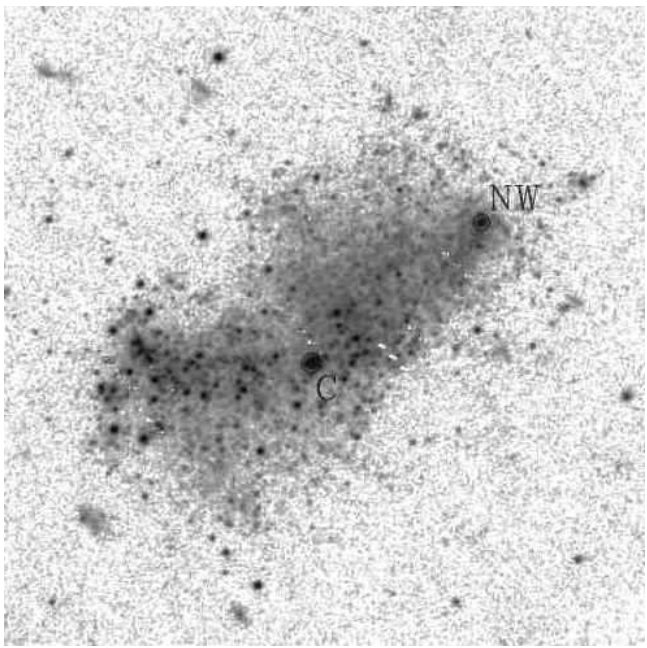}
\caption{HST/ACS image of I~Zw18C through the F555W filter. The
two clusters ``C" and ``NW" mentioned by \cite{IT04} are
reported. The image display scale is logarithmic. North is up and East
to the left. The ``white pixels" are contaminated pixels that were
rejected when performing the photometric measurements.}
\label{fig:1}
\end{figure}

\begin{figure}
\includegraphics[width=8.5cm]{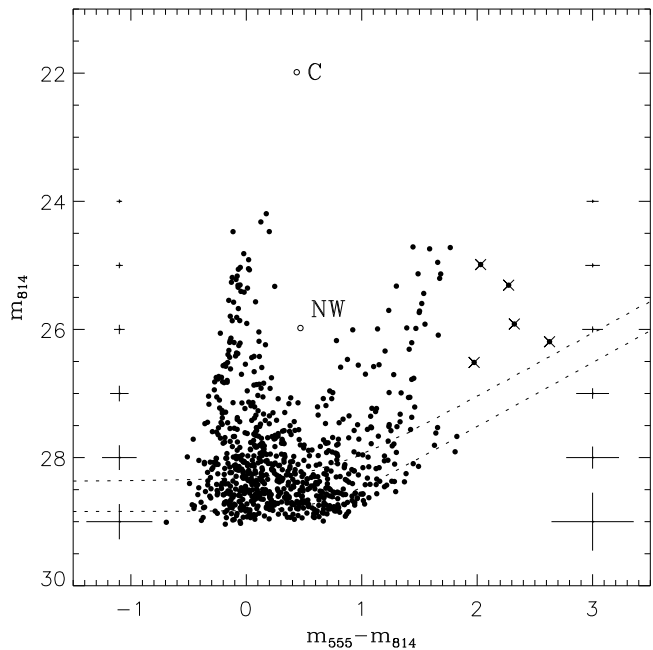}
\caption{
Color-magnitude diagram of I~Zw18C. The typical error bars as
a function of F814W are reported for both the blue and red branches.
The two clusters are annotated with open circles. The stars marked
with crosses are those we rejected from the useful sample due to their
extreme red color.We also report the 90\% and 50\% completeness limits
as dashed lines.
}
\label{fig:2}
\end{figure}

We carried out photometric measurements of the HST/ACS
images obtained in the bands F555W ($\sim$ V; Fig. \ref{fig:1}) and F814W
($\sim$I) by \cite{IT04}. They performed photometric
analysis of those images, obtaining data significantly deeper
than those of previous works. However, their procedure, 
made use of a model point spread function (PSF) that was not
optimal. For this reason, we performed our own measurements
by using the well-suited DOLPHOT/ACS package \citep{Dol00}, which correctly accounts for the pixel undersampling of
the instruments and uses accurate model PSFs.

The CMD that we obtained is
shown in Fig. \ref{fig:2}. The F555W and F814W magnitudes (hereafter
$m_{555}$ and $m_{814}$, respectively) are expressed in the VEGAMAG
system\footnote{See the HST/ACS zeropoints at
{\tt http://www.stsci.edu/hst/acs/analysis/zeropoints}.}. The CMD contains 965 sources going down to $m_{555} = 29.8$ and $m_{814} = 29.0$.We performed artificial star tests (ASTs)
to assess the quality of the photometric data and evaluate
their completeness. The latter shows a significant improvement
with respect to the measurements of \cite{IT04}. Not only is the 50\% completeness limit improved by
nearly 1 mag, but the completeness is also maintained high
closer to this limit (the completeness drop is more ``abrupt").
Furthermore, we used the ASTs to determine photometric errors
that are more realistic than the default outputs by DOLPHOT/ACS
and to evaluate the possible biases in the magnitudes retrieved.

We checked whether the stellar fluxes suffer from foreground
Galactic extinction. Whereas the survey of \cite{Setal98} yields a Galactic contribution of $A_V = 0.11$ to the extinction
toward I~Zw18, the lower-resolution data of \cite{BH82} indicate an extinction as small as $A_V = 0.01$.
Furthermore, an analysis of the nebular H$\alpha$/H$\beta$ line carried
out by \cite{Cetal02} shows that some regions of the
main component of I~Zw18 suffer very small extinction, if any.
Consequently, we assumed that the Galactic extinction toward
I~Zw18C may be negligible and we did not correct our data for
it.

\subsection{Selection of the stars}

Before carrying out the analysis of I~Zw18C, we selected the
stars for use. First, we limited the selection to the
stars situated above the 90\% completeness level. This threshold
was chosen for two reasons: (i) the uncertainties in the photometric
analysis (systematic and random errors in the magnitudes,
evaluation of the completeness map) are small for the
region of the CMD selected, and (ii) most of the stars situated
below the 90\% completeness limit are main-sequence (MS)
objects that yield little useful information about the age and
distance of the galaxy, compared with post-MS stars. Then,
we rejected the sources situated within the two clusters annotated
as ``C" and ``NW" by \cite{IT04}. Finally, we
removed the five reddest remaining stars from the selection.
According to \cite{MG07}, who revised the evolutionary
models of asymptotic giant branch (AGB) stars, the
presence of very red stars in the CMD is possible. However,
the characteristics of these objects are still very sensitive to the
physics used to model them. In principle, this problem concerns
only extremely red stars. As a consequence, we decided
to discard the stars with colors $m_{555} -  m_{814} \gtrsim 2$. The final selection
contains 408 stars, of which 315 are located in the blue
($m_{555} - m_{814} \leq 0.6$) branch and 93 in the red one.

Given the selection procedure, the completeness function
$p_c$ for the selected stars can be written as a function of the
"real" completeness $p_c^*$ evaluated for the whole CMD:

\begin{equation}
p_c = \left\{
\begin{array}{l l}
 p_c*& \mathrm{if~~} p_c* \geq 0.9 \\
0& \mathrm{otherwise.}\\
\end{array}
\right.
\label{eq:12}
\end{equation}

\subsection{Isochrones}

An important assumption of this work regards the isochrones
that we decided to use. We adopted the evolutionary tracks of
\cite{Getal00} available for $Z$ = 0.0004 and computed the
theoretical absolute magnitudes through the ACS F555W and
F814W filters\footnote{Tables of the ACS filter throughputs are available at
{\tt http://acs.pha.jhu.edu/instrument/photometry}.}
 using the BaSeL 3.1 stellar library \cite{Wetal02}.

\section{A first model}
\label{sec:4}

\subsection{Assumptions and procedure}
\label{sec:4.1}

We proceeded to a first measurement of the distance, extinction,
and SFH of I~Zw18C. The main assumptions of the calculus
were the following:

\begin{itemize}
\item a Salpeter IMF ($\phi({\cal{M}}) \propto  {\cal{M}}^{-2.35}$) in the 0.15--120
M$_\odot$ range;
\item the extinction to be uniform over the stellar
population. Furthermore, since little work has been done
regarding the extinction in I~Zw18C \cite[see, however][]{Ietal01}, we decided to use the extinction $A_V$ as a free
parameter;
\item no constraints on the SFH;
\item no concern for the possible effects of binary stars.
\end{itemize}

We explored the values of ${\cal{H}}_\mathrm{pop}$ in the $(\Delta m, A_V)$ space.
For each examined point of this space, we computed the LFs
$L_j(\tau; \Delta m, A_V)$. Using these coefficients, we evaluated the corresponding
distribution ${\cal{N}}(\tau)$ with the algorithm of \cite{Hetal99} and derived the entropy
${\cal{H}}_\mathrm{pop}(\Delta m, A_V)$. Furthermore,
we calculated the entropies ${\cal{H}}_\mathrm{dist}(\Delta m) = \log L_\mathrm{dist}(\Delta m)$ and
${\cal{H}}_\mathrm{ext}(A_V) = \log L_\mathrm{ext}(A_V)$, where $L_\mathrm{dist}(\Delta m)$ and $L_\mathrm{ext}(A_V)$ are the
LFs associated to the distance modulus and the amount of extinction,
respectively. Those two LFs are defined by

\begin{equation}
L_\mathrm{dist}(\Delta m) = \int L_\mathrm{pop}(\Delta m, A_V) \, \mathrm{d}A_V,
\label{eq:13}
\end{equation}

\begin{equation}
L_\mathrm{ext}(A_V) = \int L_\mathrm{pop}(\Delta m, A_V) \, \mathrm{d}\Delta m.
\label{eq:14}
\end{equation}

\subsection{Results}
\label{sec:4.2}

\begin{figure}
\includegraphics[width=8.5cm]{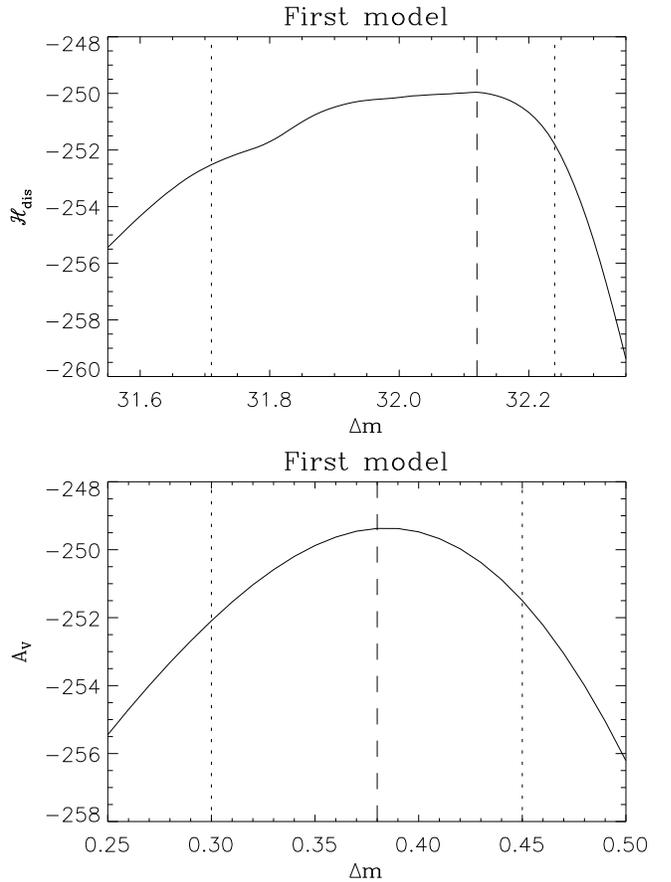}
\caption{${\cal{H}}_\mathrm{dist}$ (upper panel) and ${\cal{H}}_\mathrm{ext}$ (lower panel) curves for the model
of Sect. \ref{sec:4}. The dashed lines indicate the best-fit values of $\Delta m$ and $A_V$,
while the dotted lines delineate the 99.9\% confidence intervals.}
\label{fig:3}
\end{figure}

\begin{figure}
\includegraphics[width=8.5cm]{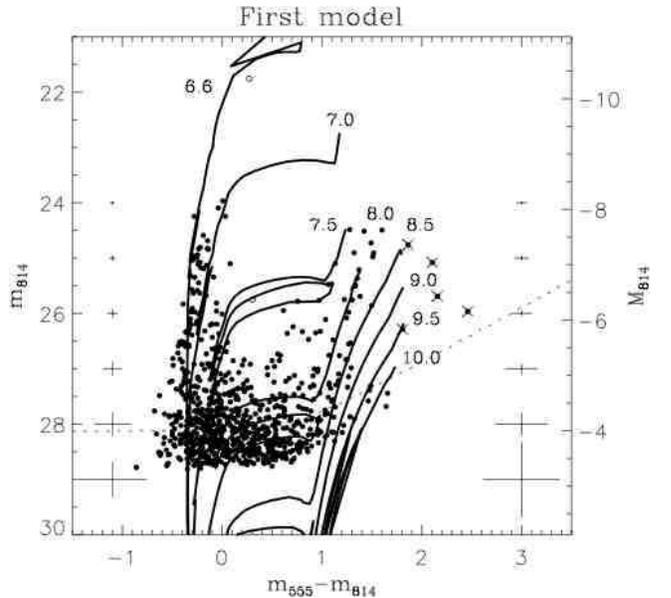}
\caption{Observed CMD and theoretical isochrones for the model of
Sect. \ref{sec:4}. The observed data are corrected for interstellar extinction, using
the best-fit value $A_V = 0.38$. The dotted line shows the threshold
above which we selected the stars to work with. The vertical axes
show both apparent and absolute F814W magnitude scales, adopting
the best-fit distance modulus $\Delta m = 32.12$.}
\label{fig:4}
\end{figure}

\begin{figure}
\includegraphics[width=8.5cm]{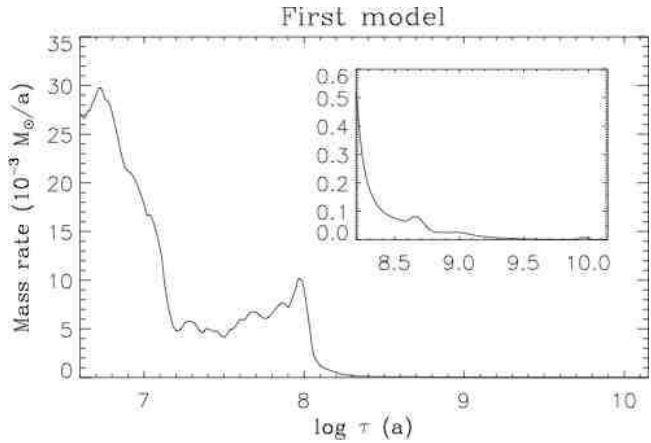}
\caption{SFH derived from the approached explained in Sect. \ref{sec:4}. The
curve was slightly smoothed to limit the apparent statistic
noise.}
\label{fig:5}
\end{figure}

The results are presented as follows. In Fig. \ref{fig:3}, we present the
curves of ${\cal{H}}_\mathrm{dist}(\Delta m)$ and ${\cal{H}}_\mathrm{ext}(A_V)$. Figure \ref{fig:4} shows the observed
CMD along with some theoretical isochrones for the best-fit
distance modulus and extinction. In Fig. \ref{fig:5}, we show the mass
rate $\dot{\cal{M}}(\tau)$ derived from ${\cal{N}}(\tau)$ at the best-fit distance and extinction.

The best-fit distance modulus is $\Delta m = 32.12$ (corresponding
to a distance of 26.5 Mpc), and the 99.9\% confidence interval
for $\Delta m$ is 31.71--32.24. Such a distance modulus is higher
than any previous estimate, by $>$ 1 mag for most of them and
$\approx$ 0.8 mag for the newest \cite[31.4,][]{Aetal07}. As for the
extinction, the best-fit coefficient is $A_V = 0.38$, and the 99.9\%
confidence interval is 0.30--0.45, a higher value than the different
estimates of the foreground Galactic extinction (see Sect.
\ref{sec:3.1}). This value falls within the 0.20--0.65 interval derived by
\cite{Ietal01} from a spectroscopic study of I~Zw18C and
suggests the existence of dust clouds responsible for extinction
inside this object.

The estimate of $\dot{\cal{M}}(\tau)$ shows various interesting features:

\begin{itemize}
\item an intense episode of SF is visible at ages  $\tau \lesssim15$ Myr.
This episode has already been detected by \cite{Ietal01},
who analyzed optical spectra of I~Zw18C, and confirmed
by \cite{IT04} in their study of the resolved
stellar population;
\item a lesser starburst is seen at  $\tau \sim 100$ Myr;
\item  approximately in the 15--70 Myr range
the SFR is comparatively moderate but not negligible between those two bursts.
Interestingly, the corresponding isochrones are those
that fall between the two branches of the CMD, although
they also cover populated regions of the CMD; that is,
this empty zone of the CMD results not from an interruption
in the SFH, but from the random filling of the CMD.
Considering the estimated SFR, an average of 4 stars would
be expected  between the branches. The likelihood
of observing no star there is about 6\%, which is not
small enough to discard the model;
\item the estimate of $\dot{\cal{M}}(\tau)$ suggests  a low-rate,
continuous SF that began at least a few hundreds Myr
ago. However, the low number of stars benefit from the
300 Myr isochrone, and the uncertainty in their ages makes
it impossible to confirm whether this SF process did occur.
\end{itemize}

Although this method seemed to yield accurate measurements
of the distance, extinction, and SFH of I~Zw18C, it
was important to assess its reliability. Because of the complexity of
the process, this could only be achieved by means of Monte-Carlo simulations. In the next section, we describe how we carried
out this test.

\subsection{Comparison to simulations}
\label{sec:4.3}

\begin{figure}
\includegraphics[width=8.5cm]{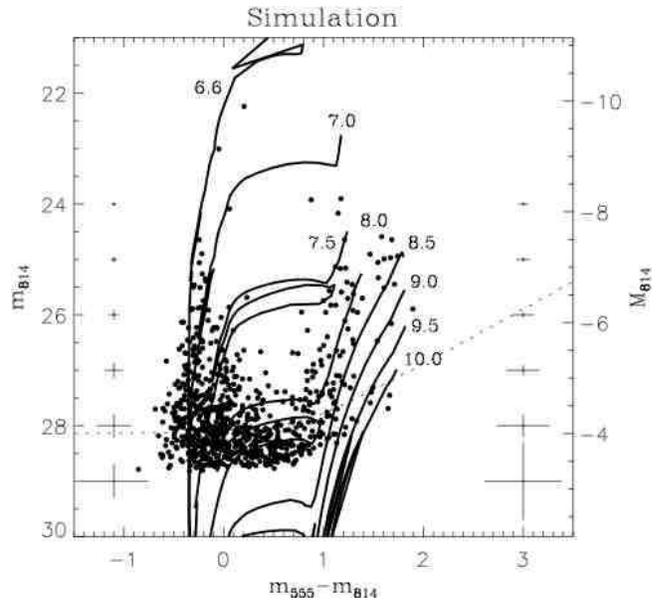}
\caption{Synthetic CMD and theoretical isochrones for one of the simulations Sect. \ref{sec:4.3}. The stars situated below the dotted line belong to
the original observations of I~Zw18C, whereas the others were simulated.
}
\label{fig:6}
\end{figure}

\begin{figure}
\includegraphics[width=8.5cm]{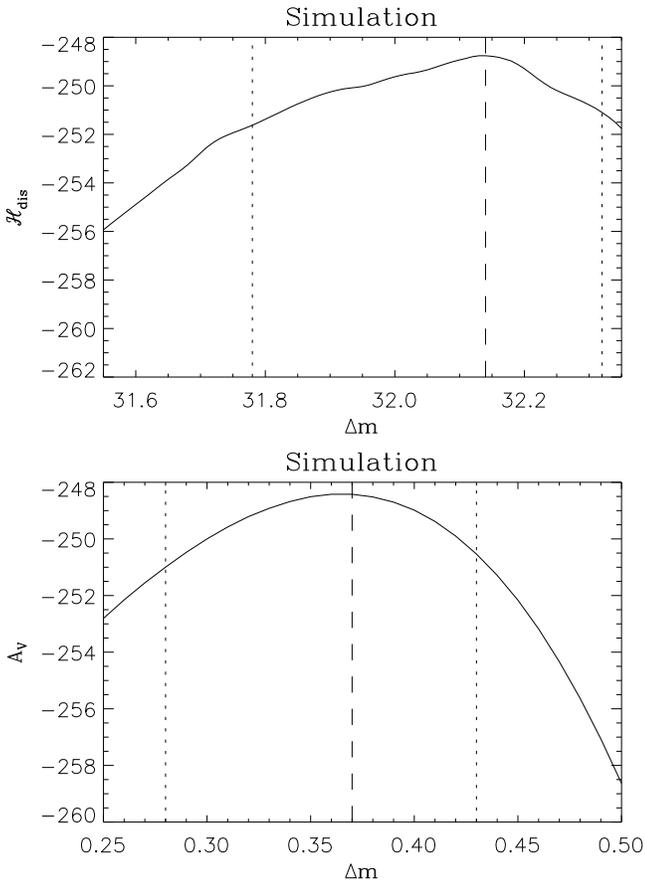}
\caption{${\cal{H}}_\mathrm{dist}$ (upper panel) and ${\cal{H}}_\mathrm{ext}$ (lower panel) curves for one of the
simulations in Sect. \ref{sec:4.3}. The dashed line indicates the best-fit value of $\Delta m$, while the dotted lines delineate the 99.9\% confidence interval.
}
\label{fig:7}
\end{figure}

\begin{figure}
\includegraphics[width=8.5cm]{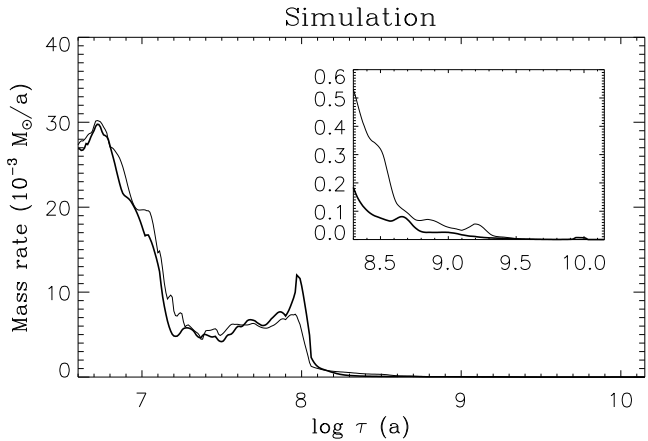}
\caption{SFH derived for one of the simulations in Sect.\ref{sec:4.3} (bold
curve). The thin curve represents the SHF estimated for I~Zw18C in
Sect. \ref{sec:4} and is shown for comparison.
}
\label{fig:8}
\end{figure}

\begin{figure}
\includegraphics[width=8.5cm]{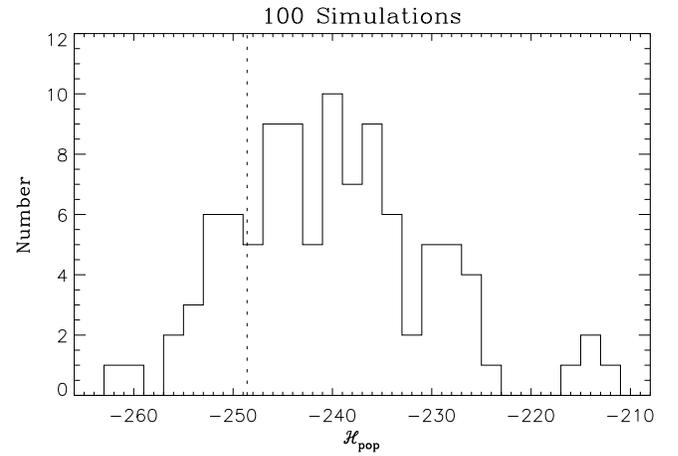}
\caption{Histogram of ${\cal{H}}_\mathrm{pop}$ for 100 simulations as described in Sect.
\ref{sec:4.3}. The superimposed curve is Gaussian and has the same mean and
standard deviation as the distribution of ${\cal{H}}_\mathrm{pop}$.}
\label{fig:9}
\end{figure}

\begin{figure}
\includegraphics[width=8.5cm]{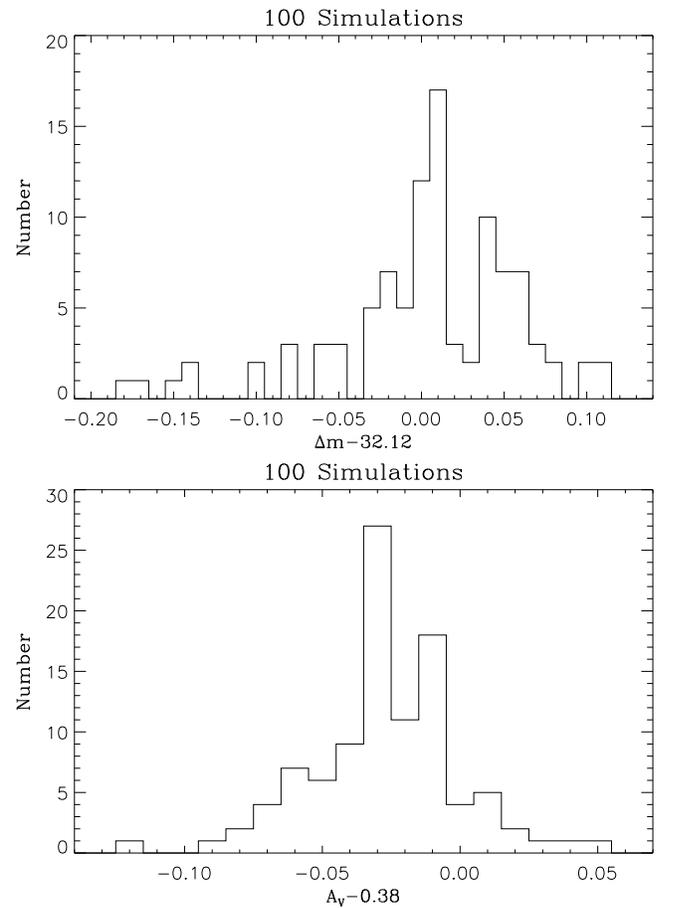}
\caption{Histograms of $\Delta m - 31.12$ and $A_V - 0.38$ for 100 simulations
as described in Sect. \ref{sec:4.3}.
}
\label{fig:10}
\end{figure}

We ran a series of Monte-Carlo simulations of stellar populations,
using the same assumptions as in Sect. \ref{sec:4.1} and the estimates
of the distance, extinction, and SFH described in Sect.
 \ref{sec:4.2}. For each simulation, we drew 408 stars with random ages
and masses (using the SFH and the IMF as distribution functions)
and photometric completeness above the 90\% threshold.

We first analyzed 3 simulated populations using a large grid
of points in the $(\Delta m, A_V)$ space to check the general
behavior of our algorithm. The three simulations yielded very
similar results. In Fig. \ref{fig:6}, we present one of the synthetic CMDs,
Fig. \ref{fig:7} shows the corresponding ${\cal{H}}_\mathrm{dist}$ and ${\cal{H}}_\mathrm{ext}$ curves, and Fig. \ref{fig:8}
 shows the mass rate computed at the best-fit distance and
extinction. There is no significant discrepancy between the distance,
extinction and SFH estimates obtained for I~Zw18C and
for the 3 simulations. In principle, this indicates that the probabilistic
method we used suffers from no severe bias.

To check for small biases in our
approach and verify the confidence level of our first model, we
performed 100 new simulations, this time exploring a limited
area of the $(\Delta m, A_V)$ space around the input values. The histograms
of ${\cal{H}}_\mathrm{pop}, \Delta m - 32.12$ and $A_V - 0.38$ are shown in Figs.
\ref{fig:9} and \ref{fig:10}. They show a possible bias in the estimation of $\Delta m$
and $A_V$, the first quantity being slightly overestimated and the
second underestimated. However, the biases encountered
are small ($\sim$ 0.03 mag) and do not exceed the respective dispersions
of the simulations. As for the ${\cal{H}}_\mathrm{pop}$ distribution, the value
obtained for I~Zw18C falls well within the range of simulated
values, an indication that the observations are well-fitted by the
model shown in this section.

Although we have been able to satisfactorily describe  the
CMD of I~Zw18C, it is important to review the assumptions so far. It is possible that modifying or relaxing
some of them results in other satisfactory models of this stellar
population.We discuss these modifications in the next section.

\section{Towards a better model for I~Zw18C}
\label{sec:5}

\begin{figure}
\includegraphics[width=8.5cm]{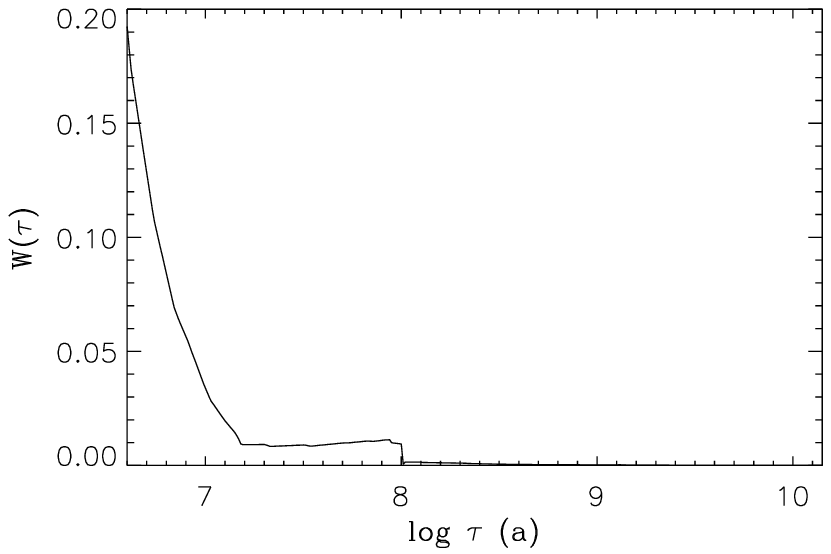}
\caption{Function $W(\tau)$ as described in Sect. \ref{sec:5.1}.}
\label{fig:11}
\end{figure}

There are at least four possible sources of error in our modeling
of I~Zw18C: the isochrones (via the evolutionary tracks, and
the model atmospheres), the IMF, the treatment of the extinction
and overlooking binarity. 

\subsection{Effects of the isochrones/stellar luminosity functions}

As explained in Sect. \ref{sec:2.2}, our method is related not just  with the isochrones used, but also with the stellar luminosity function they produce:  not only is the shape of the isochrone relevant, also the IMF (for MS stars) and the lifetime of different evolutionary phases (for post-MC stars)\footnote{Note that lifetimes of fast evolutionary phases and the amount of stars in a given evolutionary phase are linked  by the {\it fuel consumption theorem} \citep{RB84,RB86,Buzz89}; see also \cite{MG01} and \cite{CL05} for the issues about obtain isochrones from evolutionary tracks and its relation with the stellar luminosity function.}. In this way, the method provides additional constraint to the ages and distances obtained, as we see in  Sect. \ref{sec:6}. In this situation, just the shape of the isochrones from different authors is not enough to assess the uncertainties of the isochrones used.

However, we can obtain a broad idea of how the results would differ by comparing just the mean value of the stellar luminosity function produced by different isochrones. By comparing of the results of evolutionary synthesis models \cite[whose principal output is actually the mean value of the stellar luminosity function,][]{CL06}. Such a comparison has been done by \cite{Buzz05}, who in Fig. 1, shows the comparison of the evolution of U, V, K and bolometric luminosity for different synthesis models (which use different evolutionary tracks). In general, there is  very good agreement between the different synthesis models, hence evolutionary tracks, except in the K band (which strongly depends on the synthesis model implementation of AGB, and post-AGB stars). For the given photometric bands we used, it is expected that using a different set of isochrones produces (at least in a global average) similar results.

\subsection{Effects of the IMF}
\label{sec:5.1}

Changing the IMF shape modifies the values of the LFs $L_j$
through Eq. \ref{eq:2}. At a given age, this will be important if the
minimum and maximum initial masses ``visible" in the photometric
data, respectively ${\cal{M}}_\mathrm{min}(\tau)$ and ${\cal{M}}_\mathrm{max}(\tau)$, are significantly
different from each other. Let us examine the case where
the IMF is a power law of slope $\alpha$. The ratio between the LFs
$L_j(\tau)$ of a ${\cal{M}}_\mathrm{max}(\tau)$ mass star and a ${\cal{M}}_\mathrm{min}(\tau)$  one is nearly proportional
to $({\cal{M}}_\mathrm{max}(\tau)/{\cal{M}}_\mathrm{min}(\tau) )^\alpha$. As a result, we can evaluate the
impact of an uncertainty of $\Delta \alpha$ in the IMF slope on ${\cal{H}}_j(\tau)$ with
the following function:

\begin{equation}
W(\tau) = \Delta \alpha \,\log \left(\frac{{\cal{M}}_\mathrm{max}(\tau)}{{\cal{M}}_\mathrm{min}(\tau)}\right).
\label{eq:15}
\end{equation}

Even though the Salpeter IMF ($\alpha = -2.35$) is widely employed
in the study of stellar populations, various authors have
suggested the use of other IMFs. For example, according to
\cite{Sca98}, the IMF slope may vary from $\alpha = -2.2$ to 
$\alpha = -2.8$ in the ${\cal{M}} \geq 1$ M$_\odot$ range. Thus, we computed $W(\tau)$
using $\Delta \alpha = 0.5$ to check the importance of our choice of a
Salpeter IMF. The curve is shown in Fig. \ref{fig:11}. As seen,
at any age (in particular at  $\tau \geq 10$ Myr), the value of $W(\tau)$ is
low. This reflects that our star sample is almost exclusively
composed of post-MS objects whose properties,
at a given age, are very sensitive to their initial masses,
so the interval of initial masses they cover at this age is necessarily
narrow. The consequence is that the exact choice of the
IMF has no significant consequence on the calculation of the
entropies ${\cal{H}}_j(\tau)$, and changing the IMF will have little impact
on the model described in Sect. \ref{sec:4}.

\subsection{Effects of binary stars}

Our results may be affected if a significant fraction of the point
sources used are actually binary stars (either physical or optical)
where the luminosities of both components are comparable.
Assume that a given point source is a binary
of integrated magnitude $m_{814}$, with the magnitudes of the primary
and secondary component $m_{814,1}$ and $m_{814,2}$, respectively
($m_{814,2} >  m_{814,1}$). The magnitude of the secondary companion
and the ``magnitude excess" $\delta m_{814} = m_{814,1} - m_{814}$  caused by
its presence are related by

\begin{equation}
m_{814,2} = m_{814} -  2.5\, log (1 - 10^{-0.4 \delta m_{814}}).
\label{eq:16}
\end{equation}

\noindent The magnitude excess cannot be greater than
$2.5 \log 2 = 0.753$ because we impose the condition that the secondary companion be fainter than the primary one.

In spite of various attempts, it has not been clearly demonstrated
that the component masses in physical binaries are correlated
\citep{KTG93}, hence, we assumed that in a given
binary star, the component masses are independently drawn
from the IMF. For each point source, we computed the probability
$P_\mathrm{bin}$ of the magnitude excess as $\delta m_{814} \geq \sigma_{814}$, where $\sigma_{814}$ is the photometric error on $m_{814}$. We calculated this probability
assuming the source to be either a physical binary (the
two stars have the same age) or an optical double star (the ages
of the two components are independent). For all the sources of
our sample, we found $P_\mathrm{bin} <  0.15$\%, which means that binarity
is very unlikely to affect  our results seriously. This is actually
consistent with  only looking at the very brightest
stars of I~Zw18C: the chance for a random star of the galaxy to
be nearly as bright as them is small.

\subsection{Effects of spacial variations in the intrinsic extinction}

An arguable simplification of our first model is that we assumed
the interstellar extinction is uniform. In principle,
this is not the case since dust clouds may be present inside
I~Zw18C. The presence of such clouds is supported by the estimate
of the extinction in our first model, and can cause the extinction
to vary along the projected view of the galaxy. Large-scale
fluctuations may exist, depending on the morphology of
the dust clouds. Moreover, even on local scales, the extinction
may be distributed, given that the stars can be located at different
depths inside the clouds.

To our knowledge, the only work in which a heterogeneous
extinction has been measured in I~Zw18C is that of \cite{Ietal01}. They find the value of $A_V$ to range between 0.20
and 0.65. This supports the idea that treating the extinction with
some detail would be useful for more exhaustive modeling of
I~Zw18C.

\section{A model with detailed treatment of the extinction}
\label{sec:6}

\begin{figure}
\includegraphics[width=8.5cm]{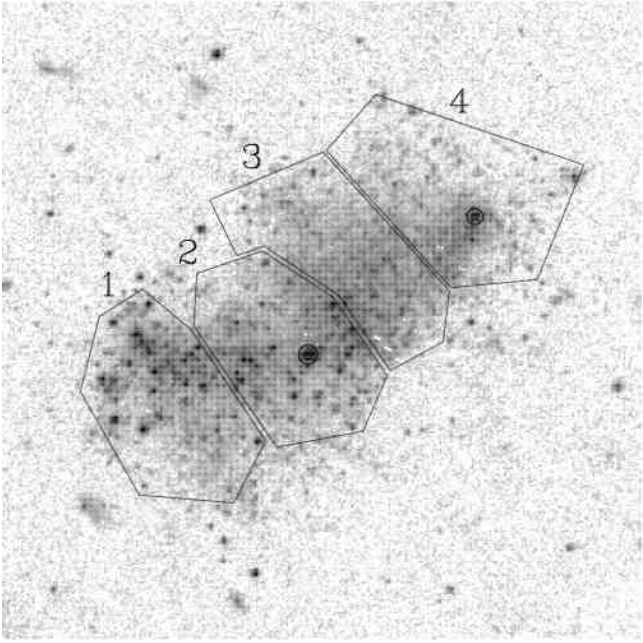}
\caption{HST/ACS image of I~Zw18C through the F555W filter, with
the boundaries of the 4 regions used in Sect. \ref{sec:6} overplotted. The two
unresolved clusters are also reported.}
\label{fig:12}
\end{figure}

We carried out a second model of the stellar content of
I~Zw18C, based on the same assumptions and procedure as the
first model, but chaning the treatment of extinction which
was dealt with in the following fashion. We divided the galaxy
into 4 regions, shown in Fig. \ref{fig:12}, that contain 101, 121, 92, and
59 stars of the original selection, respectively. In each region
$k$, we supposed the extinction of a given star is a random
variable following a uniform probability law, with $A_V$ falling in
the range $\bar{A}_V^{(k)} \pm \Delta A_V^{(k)}$.
 Mathematically, this was achieved by
replacing $L_j(\tau; Z, d, A)$ by

\begin{equation}
\begin{array}{rl}
L_j''(\tau; Z, \Delta m,  & \bar{A}_V^{(k)},  \Delta A_V^{(k)}) =  \\
& \frac{1}{2\,  \Delta A_V^{(k)}} \int_{\bar{A}_V^{(k)} - \Delta A_V^{(k)}}^{\bar{A}_V^{(k)} + \Delta A_V^{(k)}} L_j''(\tau; Z, \Delta m, A_v)\, \mathrm{d} A_V
\end{array}
\label{eq:17}
\end{equation}

We determined the best-fit values of $\bar{A}_V^{(k)}$, $\Delta A_V^{(k)}$,
and $\Delta m$ in the
following way. For given values of $\Delta A_V^{(k)}$, we computed the
${\cal{H}}_\mathrm{dist}^{(k)}(\Delta m)$ curves, derived the ${\cal{H}}_\mathrm{dist}(\Delta m) = \sum_k {\cal{H}}_\mathrm{dist}^{(k)} (\Delta m)$ distribution,
and picked the best-fit distance modulus $\Delta m$. At this
distance modulus, we computed the maximum entropies ${\cal{H}}^{(k)}$
as a function of  $\bar{A}_V^{(k)}$. We repeated this operation for different
sets of $\Delta A_V^{(k)}$ so as to maximize the values of ${\cal{H}}^{(k)}$.

\subsection{Results}
\label{sec:6.1}

\begin{table}
\caption{Best-fit values of $\Delta A_V^{(k)}$ and  $\bar{A}_V^{(k)}$ for the three regions described
in Sect. \ref{sec:6}. }
\label{tab:1}
\begin{center}
\begin{tabular}{cccc}
\hline \hline
Region & $\Delta A_V^{(k)}$ & $\bar{A}_V^{(k)}$ & (99.9\% CL)$^{\mathrm{a}}$ \\
\hline
1 & 0.10 & 0.42 & (0.25--0.50)\\
2 & 0.00 & 0.34 & (0.08--0.44)\\
3 & 0.15 & 0.32 & (0.15--0.51)\\
4 & 0.00 & 0.45 & (0.10--0.68)\\
\hline
\end{tabular}
\end{center}
\begin{list}{}{}
\item[$^{\mathrm{a}}$]The values between parenthesis give the limits of the 99.9\%
confidence intervals for  $\bar{A}_V^{(k)}$.
\end{list}
\end{table}

\begin{figure}
\includegraphics[width=8.5cm]{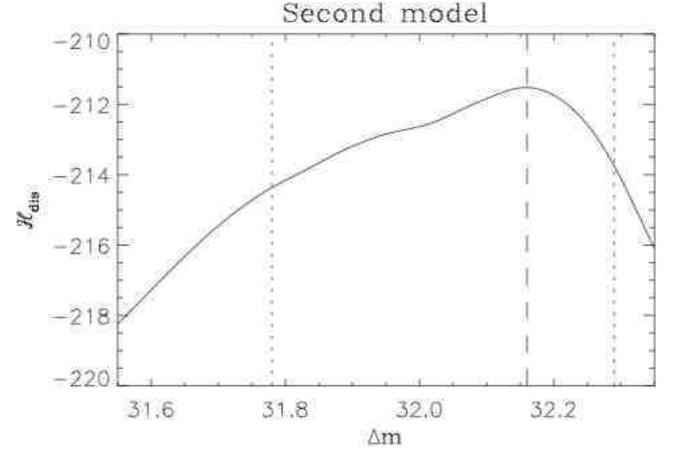}
\caption{${\cal{H}}_\mathrm{dist}$ curve for the model of Sect. \ref{sec:6}. The dashed line indicates
the best-fit value of $\Delta m$, while the dotted lines delineate the 99.9\%
confidence interval.}
\label{fig:13}
\end{figure}

The results are presented as follows. In Fig. \ref{fig:13}, we show the
curve ${\cal{H}}_\mathrm{dist}^{(k)}(\Delta m)$. The best-fit distance modulus is $\Delta m = 32.16$
(corresponding to a distance of 27.0Mpc) and the 99.9\%confidence
interval is 31.78--32.29.We found that the best-fit value
of $\Delta m$ is virtually insensitive to the adopted values of $\bar{A}_V^{(k)}$
is actually close to that found with our first model. In Table \ref{tab:1}, we summarize the values of $\bar{A}_V^{(k)}$ and $\Delta A_V^{(k)}$. All four extinction
averages $\bar{A}_V^{(k)}$
fall within the 0.20--0.65 range of \cite{Ietal01}. Furthermore, the extinction spreads $\Delta A_V^{(k)}$
represent
significant fractions of the averages for regions 1 and 3.

\subsection{The SFH of I~Zw18C and its spatial distribution}

\begin{figure}
\includegraphics[width=8.5cm]{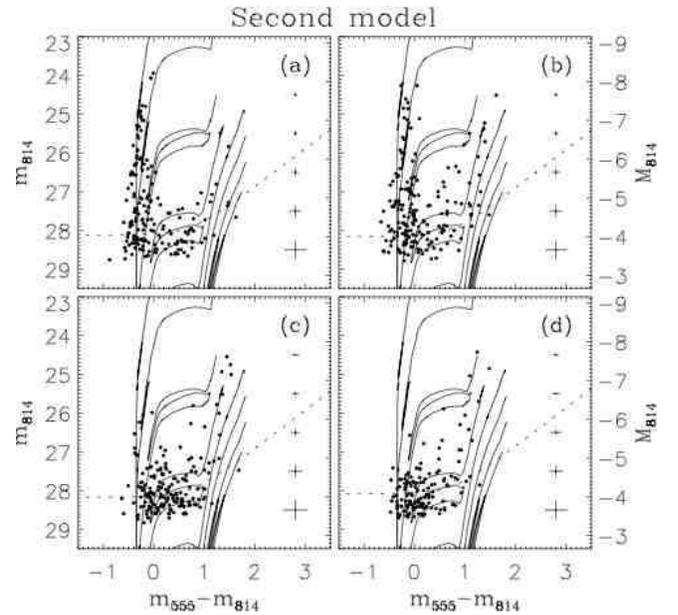}
\caption{Dereddened CMDs for (a) region 1, (b) region 2, (c) region 3,
and (d) region 4 (see Sect. \ref{sec:6}). The isochrones are the same as in Fig.
\ref{fig:4}.}
\label{fig:14}
\end{figure}

\begin{figure}
\includegraphics[width=8.5cm]{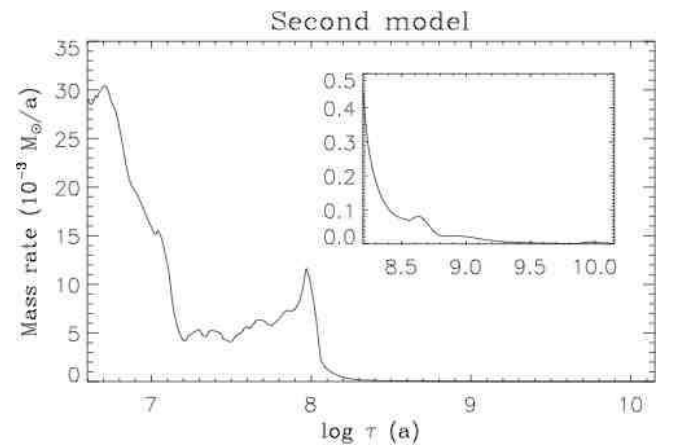}
\caption{SFH computed for region 1 (dashed line), regions 2 and 3
(dotted line), and all three regions (full line).}
\label{fig:15}
\end{figure}

\begin{figure*}
\includegraphics[width=\textwidth]{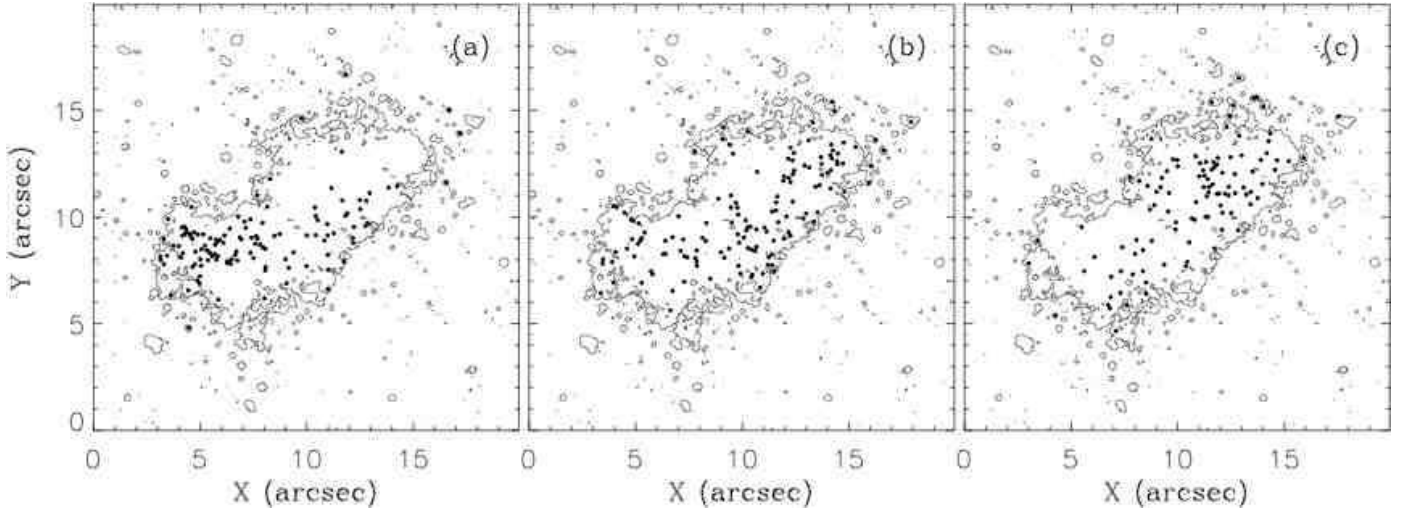}
\caption{Position of the stars falling in the following age ranges: (a) 0Ð15 Myr, (b) 15Ð60 Myr, (c) 60Ð120 Myr. For reference, a surface brightness
isocontour of the galaxy was also plotted in the panels.}
\label{fig:16}
\end{figure*}

In Fig. \ref{fig:14}, we show the CMDs for the individual regions at the
adopted distance and average extinctions. Figure \ref{fig:15} shows the
mass rate curves for the four regions together. Finally, Fig. \ref{fig:16}
shows the position of the stars in different age ranges. The age $\tau_j$ of star $j$ was defined as the age maximizing
the LF $L_j''(\tau)$.

The global SFH shown in Fig. \ref{fig:15} is nearly identical the one
discussed in Sect. \ref{sec:4}. In particular, the two starbursts at  $\tau \lesssim 15$
Myr and  $\tau \sim100$ Myr and the continuous SF between them are
confirmed with our new model. However, it is evident that the
SFH is not homogeneous across the galaxy. The young stars
($\leq 15$ Myr) tend to concentrate in the southwest tip of the object,
while there is a slight excess of 60--120 Myr old stars in the
northeast lobe. On the other hand, intermediate age stars (15--60 Myr) are distributed rather evenly through the component of
the galaxy. This observation suggests that I~Zw18C forms stars
both through a continuous, global process and through local
starbursts. The existence remains unconfirmed of the longlasting SF that started a few hundreds
to thousands of Myr.
Its possible contribution to the total mass of the
stellar population is very small, about  $10^{-3}$, and I~Zw18C
can be considered mostly as  a young object.

\subsection{The age of I~Zw18C}
\label{sec:6.3}

\begin{figure}
\includegraphics[width=8.5cm]{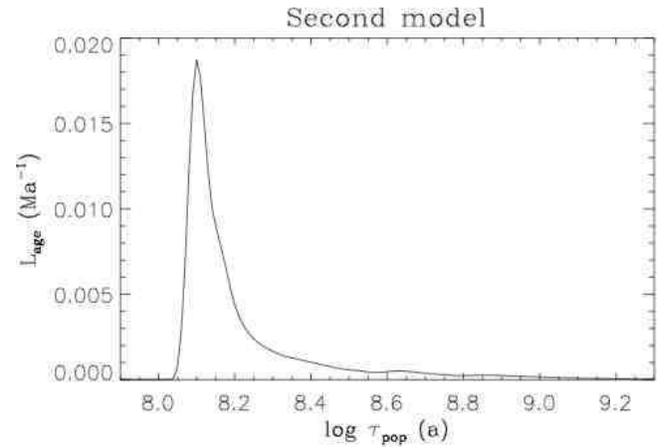}
\caption{Likelihood function of the age of I~Zw18C, as computed in
Sect. \ref{sec:6.3}.}
\label{fig:17}
\end{figure}

So far, we have presented estimates of the SFH of I~Zw18C
where the stars could be as old as 14 Gyr. These estimates do
not provide a precise constraint on the age of the stellar sample
considered, i.e. the age of the oldest of its stars. To
establish the age, we evaluated the distribution ${\cal{N}}(\tau)$  for a given tested
age $\tau_\mathrm{pop}$, with the coefficients
$G_j(\tau)$ defined as

\begin{equation}
G_j(\tau) = \int \int P(\Delta m, \bar{A}_V^{(k)})\,P_j''(\tau;\Delta m, \bar{A}_V^{(k)})\, \mathrm{d}\Delta m\,\mathrm{d} \bar{A}_V^{(k)},
\label{eq:18}
\end{equation}

\noindent forcing ${\cal{N}}(\tau) = 0$ at $\tau > \tau_\mathrm{pop}$. We then retained the corresponding entropy
${\cal{H}}_\mathrm{age}(\tau_\mathrm{pop})$ and derived the LF $L_\mathrm{age}(\tau_\mathrm{pop})$. The definition
of $G_j(\tau)$ has the advantage of  properly accounting for the probabilistic
distributions of $\Delta m$ and $\bar{A}_V^{(k)}$. With a few simulations,
we checked that this procedure yields the correct LF.

In Fig. \ref{fig:17}, we show the curve $L_\mathrm{age}(\tau_\mathrm{pop}$) obtained. If all
the stars are considered as valid elements of the sample, then
the most likely age of I~Zw18C is 125 Myr, with a 99.9\%
confidence level interval of 110 Myr--7.6 Gyr. To ensure that
($\tau_\mathrm{pop}$) $<$ 1 Gyr, it would be necessary to remove the oldest 20
stars from the sample. That is, I~Zw18C is very likely a young object,
but we cannot rule out the possibility that this object is
several Gyr old.

\subsection{A high distance estimate}
\label{sec:6.4}

\begin{figure}
\includegraphics[width=8.5cm]{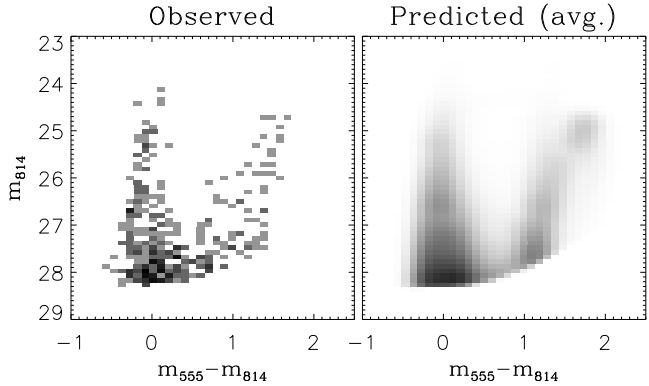}
\caption{Color and magnitude histograms of the stars considered in
Sect. \ref{sec:6}. The thick curves represent the average distributions expected
at the best-fit distance and extinction coefficients.}
\label{fig:18}
\end{figure}

\begin{figure}
\includegraphics[width=8.5cm]{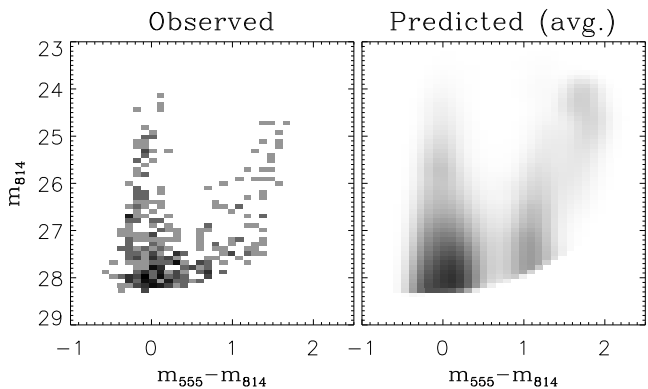}
\caption{Same as Fig. \ref{fig:18} but at a distance modulus of 31.3.}
\label{fig:19}
\end{figure}

Our estimate of the distance to I~Zw18 is by far greater than
those assumed or evaluated in all previous works published on
this object, even considering the error bars of the different estimates.
No modification to our model, including the removal of
a large number of stars from the original sample, could alleviate
this discrepancy. Whatever stars were removed, the distance
estimate was found to be very similar to that described in Sect. \ref{sec:6.1}.
A good description of the distribution of the colors and magnitudes
of the stars can be obtained by assuming the distance to
I~Zw18C to be 27 Mpc, as shown in Fig. \ref{fig:18}. In particular, the fit
of the magnitude distribution is excellent. Let us now adopt the
distance modulus of 31.3, inferred from the RBG population by \cite{Aetal07}. 
As shown in Fig. \ref{fig:19}, where the extinction
and the SFH were re-fitted, a significant excess of luminous
AGB stars ($m_{555} - m_{814} \approx 1.7$, $m_{814} \approx 24$)
is predicted, while the expected number of MS stars at the faint
end of the CMD ($m_{555} -  m_{814} \approx 0$, $m_{814} \approx 28$) is lower than
the observed one. This scheme worsens if we assume an even
shorter distance. Thus, it appears that the large distance modulus
we computed is essential for explaining the distribution of
colors and magnitudes of the stars observed.

\section{Conclusions}
\label{sec:7}

In an attempt to evaluate the distance to the dwarf galaxy
I~Zw18C and its SFH, we carried out a detailed analysis of
its resolved stellar content. We developed a probabilistic approach
based on the current knowledge about SF and its random
aspects, in order to exploit as much of the information
held by the photometric data as possible.We used high-quality
HST/ACS observations of the galaxy, on which we performed
state-of-the art photometry to obtain high signal-to-noise and
high completeness data.

We carried out a first model with a few simple assumptions
about the stellar population studied. We found the distance to
I~Zw18C to be as high as 27 Mpc and detected several important
features in the SFH: two starbursts of which an ongoing
one ($\tau \lesssim 15$ Ma) dominates the SFH and the other occurred
$\approx$ 100 Myr ago, a rather constant SF episode that has lasted for
100 Myr or so, and a possible low-rate SF process for larger
ages. We performed Monte-Carlo simulations to assess those
results. Although the simulations proved our approach to be efficient,
some of our assumptions could limit the reliability of
the model. We reviewed the assumptions and found that
only the treatment of the extinction could yield a more robust
depiction of the stellar population studied.

We performed a second model, accounting for the spatial
variations and random fluctuations of the interstellar extinction
inside I~Zw18C.We could confirm the high estimate of the distance
modulus of I~Zw18C  ($\Delta m = 32.16^{+0.13}_{-0.38}$, i.e. $d = 27.0^{+1.7}_{-4.3}$
Mpc), significantly greater than any value previously published.
The description of the SFH was also maintained. The spatial
distribution of the stars in various age intervals suggests that the
stellar content of I~Zw18C has formed both through a global,
continuous process and through a series of local starbursts.

That the distance we computed for I~Zw18C is
greater than those published in previous works is important for
study the resolved stellar population of both this object
and the main body of I~Zw18. For example, the spectra
assigned to the ionizing stars will be more intense if one adopts
our distance instead of a more conservative one. The age estimate
of the red stars is also sensitive to the adopted distance,
because it is more at shorter distances. Another consequence of our
distance estimate is that it explains why \cite{IT04}
did not detect any RGB stars in the galaxy.

An equally important result of our work regards the age of
the galaxy. Our probabilistic approach allowed us to state that
I~Zw18C is most likely young, with a maximum likelihood age
of 125 Myr, although we could not discard ages of several Gyr. A
more accurate age could be inferred from photometric data that is either
deeper or that covers a wider range of infrared wavelengths.
In any case, our estimate of the SFH shows that nearly all the
stellar content of I~Zw18C has formed in the last few hundred
Myr.

Let us finish the paper with a final note about how its results have affected the
authors of the paper (and maybe also the reader). Each of us had own expectations
about this work. Some of us expected a shorter distance for I~Zw18C and (by extension
to the main body) more consistent with pre-\cite{Ietal01} estimations: a lower distance
implies one lower absolute luminosities and a higher impact from IMF sampling effects in 
the I~Zw18 system studies. Others expected that this study would produce similar results to
the ones by \cite{IT04}.

In all cases, our preconceptions (based in different
readings and interpretations of previous works and our different personal history 
in I~Zw18 studies) have not been satisfied.

Among us, {\it a posteri} criticism has arisen (maybe without being completely aware of our 
expectations and its impact on the interpretation of the results). As an example of potential problem that the reader might also have thought about, part of our analysis is based only in a few luminous red  
and MS stars. However, our solution is based on an overall analysis of {\it both} populations
(i.e. the luminosity function) and not on a few stars in a particular box of the CMD in different areas (Figs. \ref{fig:18} and \ref{fig:19} and discussion in Sect. \ref{sec:6.4}). We are not
able to find any mistake in the methodology or in the results, and any possible criticism also applies (more strongly) to results from standard CMD methodology.

Independently of the particular expectations of each of the authors, we hope that the controversy about 
the I~Zw18 system distance and I~Zw18C star formation history will be solved with deeper observations of I~Zw18C using future large observational facilities. But for the moment, against our
expectations, it seems that current data point toward  a distance of 27 Mpc to the I~Zw18 system
and maximum likelihood age of 125 Myr.

\begin{acknowledgements}
This work was supported by the Spanish {\it Programa Nacional de  
Astronom\'\i a y Astrof\'\i sica} through FEDER funding of the project  
AYA2004-02703 and AYA2007-64712.  LJ
was supported by a UNAM post-doctoral grant. 
\end{acknowledgements}

\bibliographystyle{aa} 

\end{document}